\documentclass[%
 reprint,
showpacs,preprintnumbers,
 amsmath,amssymb,
 aps,
 pra,
 longbibliography,
 lengthcheck,%
]{revtex4-1}

\usepackage{graphicx,epstopdf, epsfig}
\usepackage{dcolumn}
\usepackage{bm}
\usepackage{hyperref}
\usepackage{units}
\usepackage{color}
\usepackage{soul}

\begin{document}

\title{Electrometry near a dielectric surface using Rydberg electromagnetically induced transparency } 
\author{R. P. Abel, C. Carr, U. Krohn, C. S. Adams}
\email{c.s.adams@dur.ac.uk}
\affiliation{Department of Physics, Durham University, Durham DH1 3LE, United Kingdom}

\date{\today}

\begin{abstract}
Electrometry near a dielectric surface is performed using Rydberg electromagnetically induced transparency (EIT). The large polarizability of high $n$-state Rydberg atoms gives this method extreme sensitivity. We show that adsorbates on the dielectric surface produce a significant electric field that responds to an applied field with a time constant of order one second. For transient applied fields (with a time scale of  $<$ 1~s) we observe good agreement with calculations based on numerical solutions of Laplace's equation using an effective dielectric constant to simulate the bulk dielectric.

\end{abstract}
\maketitle

\section{Introduction}

Highly excited Rydberg states \cite{Gallagher94} have been the subject of much recent interest with potential applications in quantum information processing \cite{Jaksch00,Lukin01,Saffman07}, non-linear optics \cite{Ashok08,Jon10}, single photon sources \cite{Walker02} and long range Rydberg molecules \cite{Pfau09}. Several recent experiments make use of Rydberg atoms that are located near to surfaces, such as the lenses used to create microscopic optical-dipole traps \cite{Saffman10,Grangier10}, the walls of thermal vapor cells \cite{Low10} or close to an atom chip \cite{BvdLvdH10}. A possible limitation of such experiments is the sensitivity of Rydberg atoms to electric fields produced by the surface, as their polarizability scales as the seventh power of the principal quantum number, $n^7$. However, this also means that Rydberg atoms are the ideal tool for probing such surface fields. 

Rydberg spectroscopy in an atomic beam has enabled background fields as small as $\pm20~\mu$V/cm \cite{Osterwalder99} to be measured using states with principal quantum number up to $n\sim500$ \cite{Neukammer87}. Similarly, spectroscopy of ultracold Rb atoms has been carried out to monitor the effect of electric fields from ions or external electrodes \cite{Bason11}. The effects of `patch' fields at conducting surfaces have been studied theoretically \cite{Martin11} and measured by studying the ionization of nearby Rydberg atoms \cite{Dunning10}. Electric fields near surfaces have also been studied using magnetically trapped Bose-Einstein condensates. The polarization \cite{Cornell04} and spatially resolved vector electric fields \cite{Cornell07} of adsorbates on metals, semiconductors and insulators have been measured. 

In this work we perform electrometry by using electromagnetically induced transparency (EIT)\cite{Fleischhauer05} to measure energy shifts of a Rydberg state \cite{Ashok07,Kev08,BvdLvdH10}. As opposed to the direct detection of Rydberg atoms by field ionization, EIT provides a non-destructive continuous probe of Rydberg level shifts without the need for additional metal components such as electrodes or ion detectors. The electrometry is carried out in close proximity to an anti-reflection coated fused-silica window and the results are sensitive to the effect of adsorbates on the surface and the polarizability of the dielectric.

The paper is organised as follows: in Sec.\ II we discuss the experimental setup. In Sec.\ III we consider a general model of EIT in the weak probe limit that is used to extract the value of the local electric field in the vicinity of the atoms. Finally in Sec.\ IV we show that adsorbates on a dielectric surface produce a significant time-dependent field for the atoms. For the regime in which this field can be neglected, we show that the bulk dielectric can be modelled using an effective dielectric constant.

\section{Experimental setup}

\begin{figure}
 \includegraphics{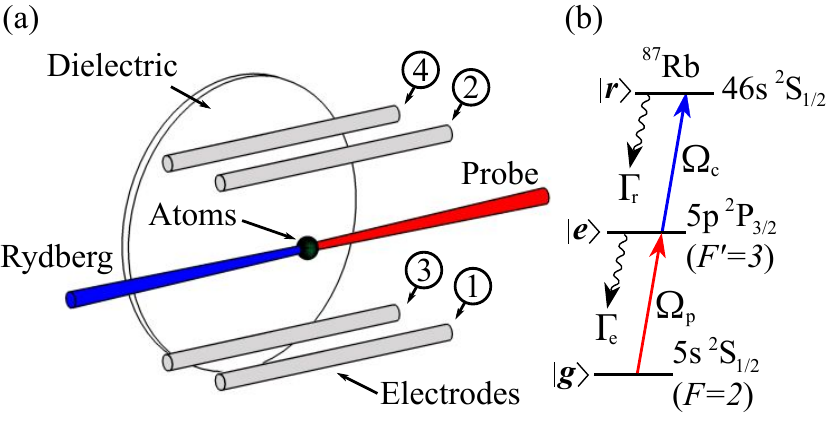} \centering
\caption{(a) Experimental setup for electrometry using Rydberg EIT. A probe beam and counter-propagating Rydberg coupling beam are focused onto an ultracold sample of $^{87}$Rb atoms released from a MOT. The probe beam is scanned through resonance and detected on a fast photodiode. An electric field is produced using four rod shaped electrodes as shown in the diagram. The measurement is carried out near a dielectric surface positioned 12.5~mm from the MOT. (b) Energy level scheme for two photon cascade system EIT in $^{87}$Rb. A probe, with Rabi frequency $\Omega_{\rm p}$, is scanned through the 5s $^2$S$_{1/2}$ $\rightarrow$ 5p $^2$P$_{3/2}$ transition which is coupled to the 46s $^2$S$_{1/2}$ Rydberg state by $\Omega_{\rm c}$. }
\label{Fig:SetupEnergyLevels}
\end{figure}

The experimental setup is shown in Fig.\ \ref{Fig:SetupEnergyLevels}(a). Electric fields are applied using four rod-shaped stainless steel electrodes, 3~mm in diameter and positioned in a rectangular pattern measuring 34~mm by 15~mm. The laser cooled atoms are at the center of the rectangle at a position that is 12.5~mm from the dielectric surface. Counter propagating probe and Rydberg coupling beams are focused at the position of the atoms to a $1/{\rm e}^2$ diameter of 100~$\mu$m. The power of the probe beam is 60~nW and the Rydberg coupling beam 45~mW. Both beams are circularly polarized to maximise transition strengths. After passing through the sample of cold atoms, the probe beam is separated from the Rydberg coupling beam using a dichroic mirror and detected on a photodiode. The probe beam is derived from a commercial 780.24~nm diode laser stabilized to the 5s~$^2$S$_{1/2}$ $(F=2)$ $\rightarrow$ 5p~$^2$P$_{3/2}$ $(F^{\prime}=3)$ transition using modulation transfer spectroscopy \cite{McCarron08}. Light for the Rydberg coupling beam is provided by a commercial frequency doubled diode laser system with wavelength 479--486~nm. The laser is stabilized to the 5p~$^2$P$_{3/2}$ (F=3) $\rightarrow$ 46s~$^2$S$_{1/2}$ transition using cascade system EIT in a room temperature, isotopically pure $^{87}$Rb vapor cell \cite{Abel08}. 

We load $10^6$ $^{87}$Rb atoms into magneto-optical trap (MOT) \cite{Chu87}. The MOT is loaded from a background vapor of Rb atoms produced by a heated dispenser. An optical molasses phase with a duration of 7~ms cools the atoms to a temperature of 30~$\mu$K. After a 5~ms period of free expansion the atoms are prepared in the $|F = 2, m_F = 2\rangle$ state with a 0.2~$\mu$s optical pumping pulse. An EIT spectrum is then measured by sweeping the probe beam over the D$_2$ $F = 2 \rightarrow F^{\prime} = 3$ transition while the Rydberg coupling beam is also on. The probe frequency sweep is completed in 250~$\mu$s and extends 20~MHz either side of resonance. A fast photodiode, with a bandwidth of 20 MHz, is used to detect the probe and an EIT spectrum is recorded.

\section{Electromagnetically induced transparency}

In this work we consider the cascade system EIT shown in the energy level diagram in Fig.\ \ref{Fig:SetupEnergyLevels}(b) where the ground state, $|g\rangle$, is coupled to an excited state, $|e\rangle$, with a probe beam of Rabi frequency $\Omega_{\rm p}$, which is in turn coupled to the Rydberg state, $|r\rangle$, by a coupling beam of Rabi frequency $\Omega_{\rm c}$. The states $|e\rangle$ and $|r\rangle$ have natural decay rates $\Gamma_{\rm e}$ and $\Gamma_{\rm r}$ respectively. The probe susceptibility \cite{Banacloche07}, in the limit where the probe is weak and the Doppler shift is negligible, is given by:

\begin {equation}
\chi_{\rm p} = \frac{{\rm i}Nd^2_{\rm ge}/\epsilon_0\hbar}{\gamma_{\rm eg}-{\rm i}\Delta_{\rm p}+\frac{ \Omega^2_{\rm c}/4}{\gamma_{\rm rg}-{\rm i}(\Delta_{\rm p}+\Delta_{\rm c})}}, 
\label{eq:susceptibility}
\end{equation}

\noindent where $N$ is the atom number density, $d_{\rm ge}$ is the dipole matrix element for the  $|g\rangle \rightarrow |e\rangle$ transition and $\Delta_{\rm p,c}$ are the detunings of the probe and coupling beams respectively. The decay rates, $\gamma_{ij}$, are related to the natural decay rates by $\gamma_{ij} = (\Gamma_i+\Gamma_j)/2$ (where $\Gamma_g = 0$). The absorption of the probe laser through a sample of atoms can be calculated from the imaginary part of the susceptibility providing an analytical function to fit the experimental data. Although values of the decay rates and Rabi frequencies only give meaningful values in the strict weak probe limit, the probe and Rydberg coupling beam detunings can still be found using this method for larger Rabi frequencies.


If the coupling laser is tuned to resonance then in an applied electric field ${\cal E}$ one can write:

\begin{equation} 
\frac{\Delta_{\rm c}}{2\pi}=-\frac{1}{2}\alpha_n{\cal E}^2,
\label{eq:Starkshift}
\end{equation}

\noindent where $\alpha_n$ is the polarizability of the Rydberg state in units of MHz/(V/cm)$^2$. As the polarizability is known ($\alpha_{46}= 28.3$~MHz/(V/cm)$^2$ for the 46S state in Rb used below \cite{Stoicheff85}), the electric field can be calibrated by applying a known voltage to one of the electrodes and measuring the resulting shift $\Delta_{\rm c}/2\pi$ of the Rydberg energy level.







\begin{figure}
 \includegraphics{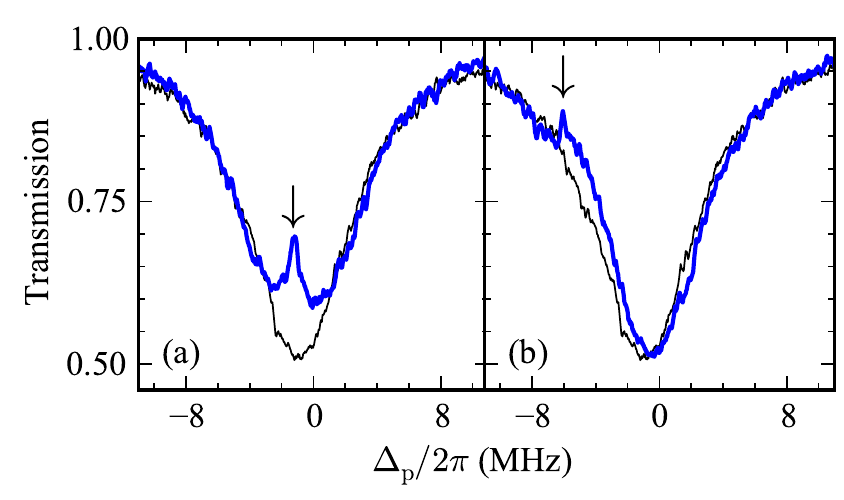} \centering
\caption{ The black lines show the transmission of a 60~nW probe beam as the detuning is scanned through resonance, in a time of 250~$\mu$s. The blue line shows the probe beam transmission in the presence of a 45~mW Rydberg coupling beam. The presence of the Rydberg coupling beam is manifest as a transparency feature, indicated by the arrows, that is superposed onto the Lorentizian absorption lineshape. In panel (a) there is no external electric field applied by the electrodes and the EIT feature appears at zero probe detuning. In panel (b) the EIT feature appears shifted due to the presence of an applied electric field.  }
\label{fig:EIT_example}
\end{figure}

In Fig.\ \ref{fig:EIT_example} we present an example of probe transmission, as it is scanned through resonance, with and without the Rydberg coupling beam (blue and black lines respectively). The peak absorption is shifted to the red due to a quantization magnetic field of 2.5~G applied parallel to the probe beam during the optical pumping and probe scan. The data showing the probe transmission in the presence of the Rydberg coupling beam displays an EIT feature. In addition to the narrow EIT peak, the probe transmission shows a broader pedestal which we attribute to optical pumping processes. In panel (a) in Fig.\ \ref{fig:EIT_example} there is no voltage applied to the electrodes so that the EIT feature appears at zero probe detuning. When a voltage is applied to the electrodes, as is the case in panel (b) of Fig.\ \ref{fig:EIT_example}, the EIT feature is observed to shift as expected. A lineshape calculated from Eq.\ \eqref{eq:susceptibility} can be used to fit these data and obtain values for the Rydberg energy level shift $\Delta_{\rm c}$ and thus the local electric field using Eq.\ \eqref{eq:Starkshift}.

\begin{figure}
 \includegraphics[]{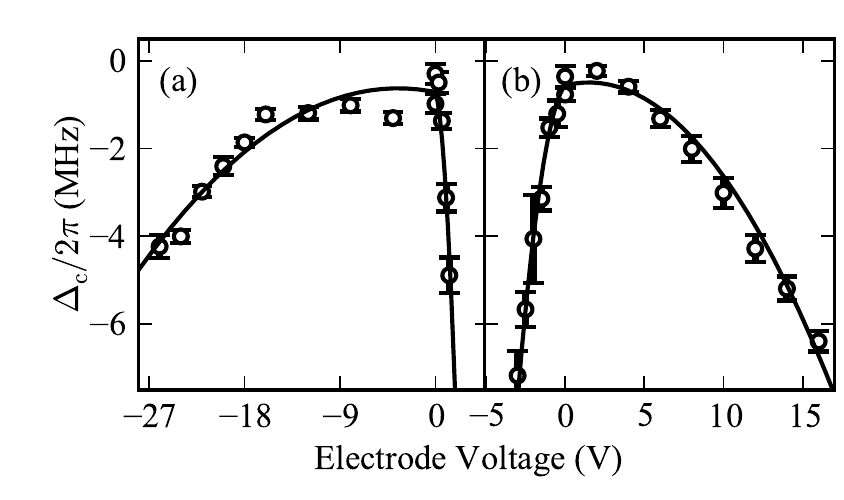} 
\centering
\caption{Energy level shift measurement of the 46S Rydberg state as a function of the voltage applied to the electrodes. The voltage is applied to the electrodes throughout the duration of the experiment. In panel (a) the voltage is applied to electrodes \textcircled{\small 3} and \textcircled{\small 4}; in panel (b) the voltage is applied to \textcircled{\small 1} and \textcircled{\small 2}. The solid line is a fit to the data using Eq.\ \eqref{eq:StarkShift}.   }     
\label{fig:CW_StarkMap}
\end{figure}

\section{Electrometry}

The measured Rydberg energy level shift as a function of the voltage applied to electrodes is presented in Fig.\ \ref{fig:CW_StarkMap}. In panel (a) the voltage is applied to the electrodes labeled \textcircled{3} and \textcircled{4}; panel (b) shows the results of applying the voltage to electrodes \textcircled{1} and \textcircled{2}. The voltage is applied to the electrodes throughout the duration of the experiment. Each data point represents the average of 10 experimental runs with the standard error indicated by the error bars. According to Eq. \eqref{eq:Starkshift} the Stark shift is expected to be quadratic in the field. However, the data display a strong asymmetry about the zero field point. The data are fit with a function of the form:    

\begin{equation} 
\frac{\Delta_{\rm c}}{2\pi}= - \frac{1}{2}\alpha_n (\mathcal{E}+\beta(\tau_{\rm p})|\mathcal{E}|+\mathcal{E}_0)^2,
\label{eq:StarkShift}
\end{equation}

\noindent where $\mathcal{E}_0$ is a constant background electric field and $\beta(\tau_{\rm p})\vert\mathcal{E}\vert$ is an additional field component which as we will show below depends on the length of time $\tau_{\rm p}$ that the voltage is applied to the electrodes. The $\beta(\tau_{\rm p})|\mathcal{E}|$ term in Eq.\ \eqref{eq:StarkShift} reproduces the observed asymmetry. 

\begin{figure}
\includegraphics[]{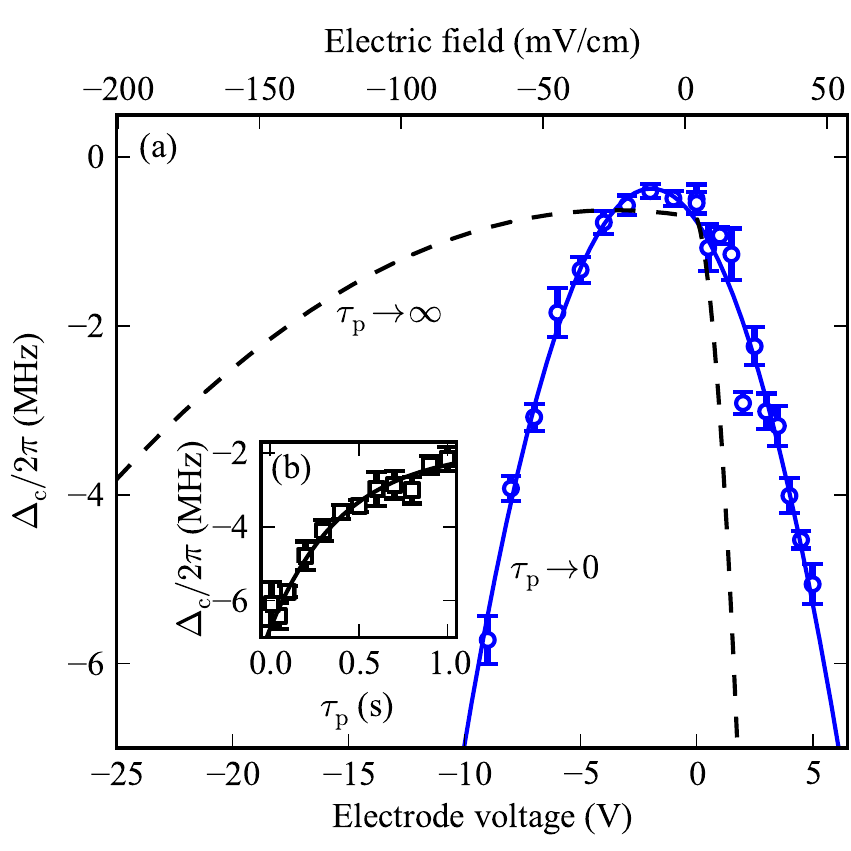} 
\centering
\caption{(a) Energy level shift of the 46S Rydberg state for contrasting pulse length regimes. The blue data are the measurement of the Rydberg energy level shift where the electrode voltage pulse length tends to zero $(\tau_{\rm p}\rightarrow 0)$. These data are fit with a quadratic (function) and the corresponding electric field found from the polarizability. The dashed line shows the Rydberg energy shift where the electrode voltage is applied continuously $(\tau_{\rm p}\rightarrow \infty )$. (b) Shows the Rydberg energy level shift as a function of the voltage pulse length with the voltage fixed at -9~V. }
\label{fig:Figure4}
\end{figure}

We attribute the electric field represented by $\beta(\tau_{\rm p})|\mathcal{E}|$ to the presence of adsorbates on the dielectric surface. In our experiment the adsorbates most likely arise from the use of alkali metal dispensers which are directed towards the dielectric. 
The electric fields produced by adsorbates on various materials have also been studied by Cornell \emph{et al.\ }\cite{Cornell04}\cite{Cornell07} by measuring their effect on a Bose-Einstein condensate held in a magnetic trap. When an atom sticks to a dielectric its charge is modified producing a dipole \cite{Cornell04}. The valence electrons of the adsorbate reside partially within the dielectric leaving a net positive charge at the surface. This results in a dipolar field that is proportional to the applied field, but always points away from the surface. When the voltage is applied to the electrodes nearest the dielectric surface ({\textcircled{\small 3} and \textcircled{\small 4} in Fig.\ \ref{Fig:SetupEnergyLevels}), fields $\mathcal{E}$ and $\beta(\tau_{\rm p})|\mathcal{E}|$ add destructively and the overall field is reduced. When the same electrodes carry a positive voltage, $\mathcal{E}$ points in the opposite direction, $\mathcal{E}$ and $\beta(\tau_{\rm p})|\mathcal{E}|$ add constructively, and the overall field is enhanced. This behaviour is confirmed by the data shown in  Fig.\ \ref{fig:CW_StarkMap}.

The measured Stark map is dramatically modified if we apply a transient electric field. Similar time dependent effects have also observed in vapor cells \cite{Ashok07} and in cold atoms \cite{Bason11}. The effect of the electrode voltage pulse length, $\tau_{\rm p}$, on the Rydberg energy level shift is presented in Fig.\ \ref{fig:Figure4}.  Two regimes are considered: the dashed line is the fit to the data shown in Fig.\ \ref{fig:CW_StarkMap}(a), where the Rydberg energy level shift is measured when the voltage is applied continuously $(\tau_{\rm p}\rightarrow \infty )$; the blue data show the Rydberg energy level shift where the voltage is pulsed ($\tau_{\rm p}\rightarrow 0$). In this case the voltage applied to the electrodes is on only for a time of 110~$\mu$s, coinciding with the range of the probe scan where the EIT feature is observed. Here the Rydberg energy level shift clearly has a quadratic dependence on the applied voltage and the data are fit accordingly with Eq.\ \eqref{eq:StarkShift} with $\beta=0$. 

\begin{figure}
\includegraphics{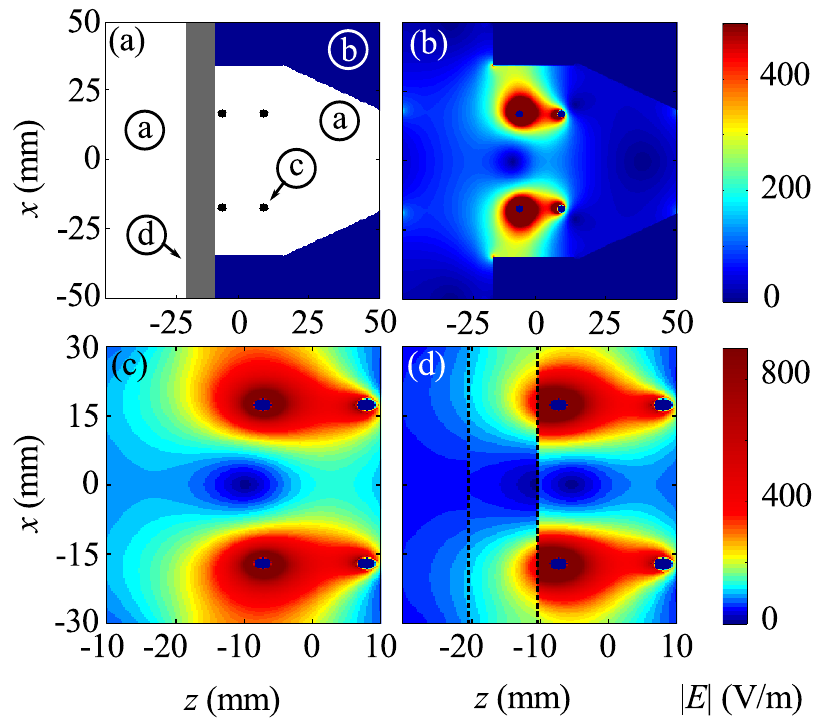} \centering
\caption{  Calculation of the electric field by numerically solving Laplace's equation. (a) Shows the geometry of the known regions of potential where: \textcircled{a} is free space; \textcircled{b} is the grounded metal bulk of the vacuum chamber; \textcircled{c} are the electrodes; and \textcircled{d} is the dielectric. (b) Shows the magnitude of the electric field without the dielectric. Parts (c) and (d) show the electric field magnitude in the region of the electrodes without and with the dielectric respectively. In (d) the boundary of the dielectric is indicated by the vertical dotted lines.       }
\label{fig:EfieldColourMap}
\end{figure}

The relaxation time between the $\tau_{\rm p}\rightarrow 0$ and $\tau_{\rm p}\rightarrow \infty$ Stark map is determined by measuring the Stark shift $\Delta_{\rm c}$ at a fixed electrode voltage of $-9$~V as a function of the duration $\tau_{\rm p}$ of the applied field. The results are  shown in Fig.\ \ref{fig:Figure4}(b). The solid line is an exponential fit of the form $1-\text{exp}(-\tau_{\rm p}t/T_{\rm a})$ with a time constant $T_{\rm a}\sim1$~s. The time $T_{\rm a}$ gives an indication of the time response of the adsorbate dipoles. It follows that the electric field contribution of adsorbate dipoles  can be  minimized by reducing the time that electric fields is applied, i.e., by using an electric field pulse with duration $\tau_{\rm p}\ll T_{\rm a}$.  


Consequently, if the applied field is pulsed, we observe the expected quadratic Stark effect as shown in Fig.\ \ref{fig:Figure4}. To check whether the observed Stark shift agrees with the expected value we calculate the applied field by numerically solving Laplace's equation in two dimensions $\nabla^2\phi(x,z) = 0$ for the exact geometry of the vacuum chamber and electrodes. The key areas included in the calculation are shown in Fig.\ \ref{fig:EfieldColourMap}(a). The areas marked \textcircled{a} are free space, \textcircled{b} is the metal vacuum chamber, which is fixed at zero potential, \textcircled{c} are the electrodes and \textcircled{d} is the dielectric surface.

\begin{figure}
\includegraphics{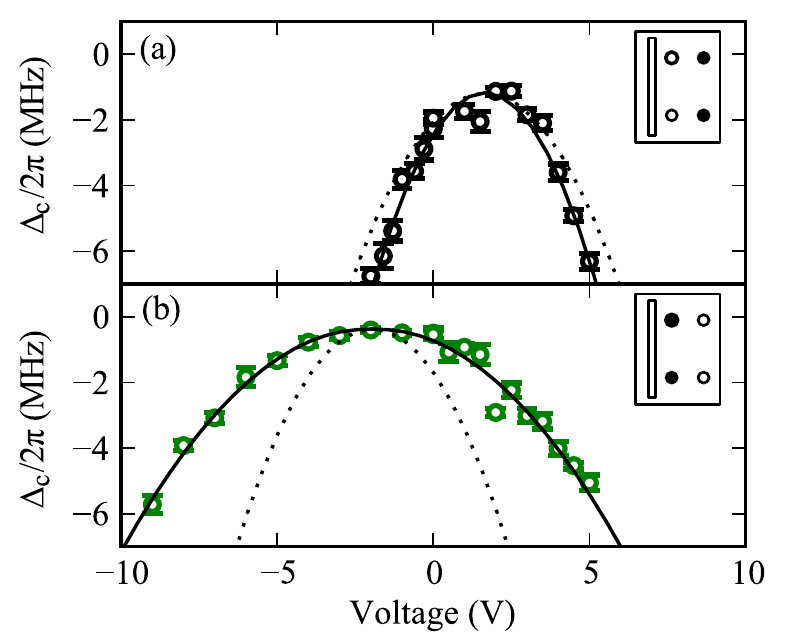} \centering
\caption{Comparison of measured Rydberg energy level shift with that corresponding to the calculated electric field. Each panel shows data for a different electrode geometry relative to the dielectric surface, where grounded electrodes are unfilled  while those with an applied voltage are filled black. The dotted lines show the results of the calculated electric field when the dielectric surface is not included in the calculation. The solid lines show the case where the dielectric is included resulting in improved agreement to the data.     }
\label{fig:StarkMap_TheoryExpt}
\end{figure}

The calculation is carried out in two parts: first the potential throughout the chamber is calculated without the dielectric. The results for an electrode potential of $-$8~V (on both electrodes nearest the dielectric and  0~V on the other two) are shown in Fig.\ \ref{fig:EfieldColourMap}(b) and (c). Secondly the calculation is repeated with the potential in the dielectric reduced by a factor $\eta$.  The value of $\eta$ is chosen such that agreement is found between the model and the measured Rydberg energy level shifts. Clearly within the bounds of the dielectric, indicated by the dotted vertical lines in Fig.\ \ref{fig:EfieldColourMap}(d), the electric field is reduced and consequently the field experienced by the atoms is also reduced. 



The agreement between the electric field calculations and the observed shifts, including the effect of the dielectric surface, are shown in Fig.\ \ref{fig:StarkMap_TheoryExpt}. The two subplots correspond to different applied voltage configurations as shown in the insets. The dotted and solid lines show the calculated shift without and with the dielectric surface respectively. For the best fit the electric field in the dielectric is reduced by a factor $\varepsilon_{\rm r}=1.6$. This value is less than the low frequency permittivity of silica glass, 3.81 \cite{CRC}, suggesting that surface charges play an additional role in determining the field.

 These results highlight the large effect of the dielectric on the electric field produced at the position of the atoms. This is most apparent where the live electrodes are in close proximity to the dielectric and in this case the electric field is seen to be reduced by 50\%.

\section{Conclusion}

We have presented a method for performing electrometry near a dielectric surface by measurement of Rydberg energy levels shift using electromagnetically induced transparency. At a modest principle quantum number of $n=46$ we demonstrate a sensitivity to background electric fields of $\sim10$~mV/cm. Where electric fields are applied continuously, the Rydberg energy level shift is found to depart from the expected quadratic dependence on applied electrode voltage. We attribute this affect to the presence of adsorbates on the dielectric surface. In the regime where the electric field is applied for a short time, a quadratic Rydberg energy level shift is recovered. By comparing the measured field to numerical solutions of Laplace's equation we have found that the dielectric contributes a field component half the magnitude of the applied field.

For Rydberg states of increasing principal quantum number the $n^7$--scaling of the polarizability for s--states, or the larger polarizability of higher angular momentum states allows this method to achieve extreme sensitivity. It may therefore find application in precision measurements of surface adsorbates and dielectric properties. It is of general importance for experiments that make use of Rydberg atoms located near dielectric surfaces where surface perturbations need to be minimal.

\section*{Acknowledgements}
We are grateful to J Millen for assistance with the electric field calculation, D P Hampshire for fruitful discussions and M P A Jones for help preparing the manuscript. We thank the EPSRC for financial support.

\end{document}